\documentclass[preprint,11pt,authoryear]{elsarticle}

\usepackage{graphicx,psfrag,color,hyperref}
\usepackage{amssymb,latexsym,mathrsfs,amsmath}
\usepackage{enumerate}

\journal{Chaos, Solitons \& Fractals}

\begin{document}

\begin{frontmatter}



\title{Branching Brownian Motion\\ Conditioned on Particle Numbers}


\author{Kabir Ramola, Satya N. Majumdar and Gr\'egory Schehr}
\address{Laboratoire de Physique Th\'{e}orique et Mod\`{e}les Statistiques, UMR 8626,\\
Universit\'{e} Paris-Sud 11 and CNRS,
B\^{a}timent 100, Orsay F-91405, France}

\begin{abstract}
We study analytically the order and gap statistics
of particles at time $t$ for the 
one dimensional branching Brownian 
motion, conditioned to have a fixed number of particles
at $t$. The dynamics of the process proceeds
in continuous time where at each time step, every particle in the system
either diffuses (with diffusion constant $D$), dies (with rate $d$) 
or splits into two independent particles (with rate $b$). 
We derive exact results for the probability distribution function of 
$g_k(t) = x_k(t) - x_{k+1}(t)$, the distance between successive particles,
conditioned on the event that there are 
exactly $n$ particles in the system at a given time $t$.
We show that at large times these conditional distributions become stationary $P(g_k, t \to \infty|n) = p(g_k|n)$.
We show that they are characterised by an exponential tail 
$p(g_k|n) \sim \exp[-\sqrt{\frac{|b - d|}{2 D}} ~g_k]$ for large gaps 
in the subcritical ($b < d$) and supercritical ($b > d$) phases, 
and a power law tail $p(g_k) \sim 8\left(\frac{D}{b}\right){g_k}^{-3}$ 
at the critical point ($b = d$), independently of $n$ and $k$.
Some of these results for the critical case were announced in a recent letter
[K. Ramola, S. N. Majumdar and G. Schehr, Phys. Rev. Lett. {\bf 112}, 210602 (2014)].
\end{abstract}


\end{frontmatter}


\section{Introduction}
\begin{figure}
\hspace{2.5cm}\includegraphics[width=0.5\linewidth]{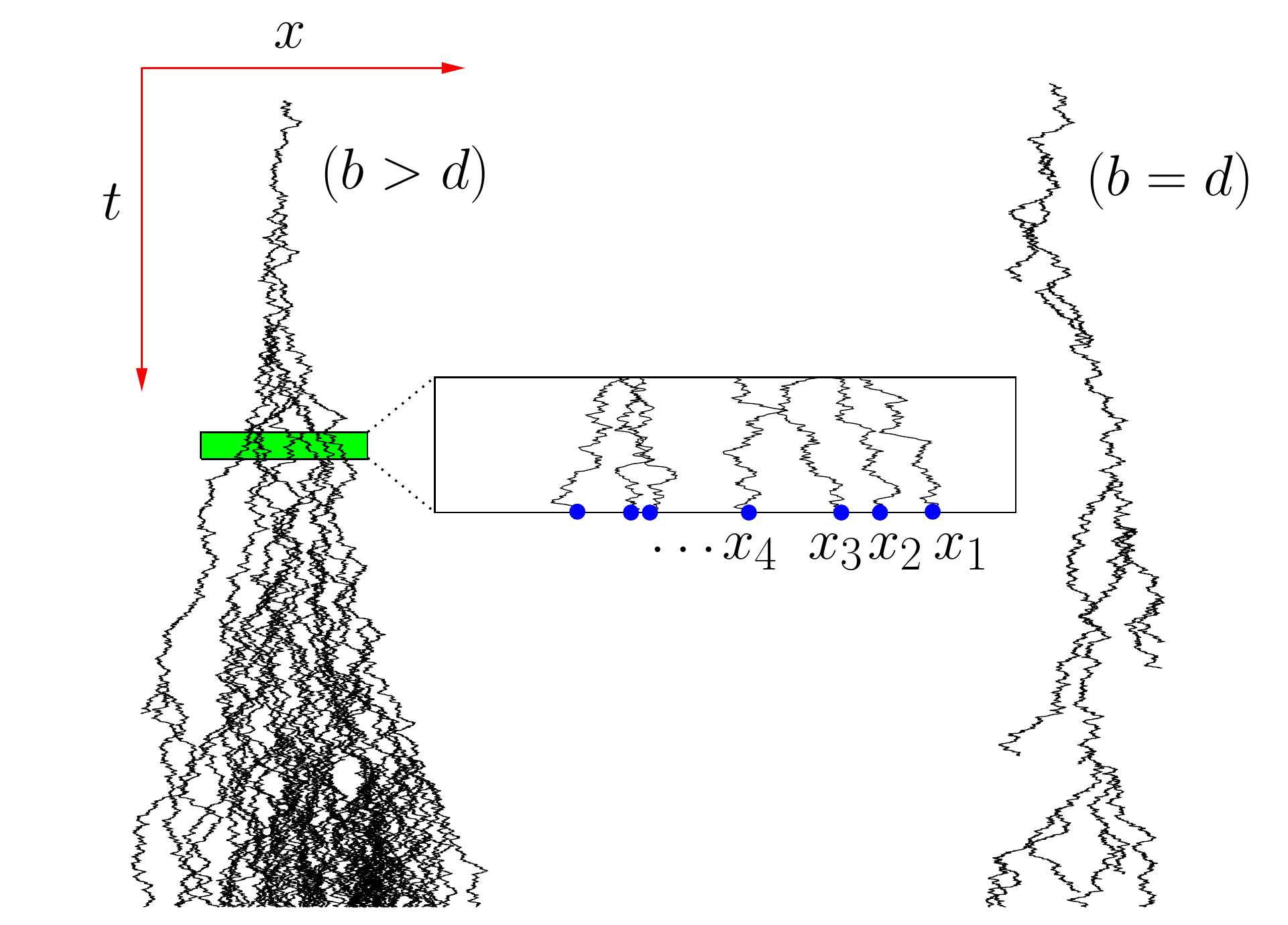}
\caption{A realization of the dynamics of branching Brownian motion with death (left) in 
the supercritical regime ($b > d$) and (right) in the critical regime ($b = d$). 
The particles are numbered sequentially from right to left
as shown in the inset.}
\label{walk_picture}
\end{figure}

Branching processes are prototypical models of systems where new particles are generated at every time 
step -- these include models of evolution, epidemic spreads and nuclear reactions amongst others
\cite{fisher,golding,derrida_brunet_simon,sawyer,majumdar_pnas}. 
An important model in this class is the Branching Brownian motion (BBM).
We focus in this paper on the simple one-dimensional BBM, where the process starts with a single particle at the origin $x = 0$
at time $t = 0$. The dynamics proceeds in continuous time according to the following rules. In a small time interval $\Delta t$, 
each particle performs one of the three following microscopic moves:  
(i) it splits into two independent particles with probability $b \Delta t$, (ii) it dies with probability $d \Delta t$ and 
(iii) with the remaining probability $1-(b+d)\Delta t$ it
performs a Brownian motion moving by a stochastic distance $\Delta x (t) = \eta(t) \Delta t$.
Here $\eta(t)$ is a Gaussian white noise with zero mean and delta-correlations with 
\begin{align}
\langle \eta(t) \rangle =0, ~~ \langle \eta(t_1)\eta(t_2) \rangle = {2D}\delta(t_1-t_2) 
\label{noise_def}
\end{align}
where $D$ is the diffusion constant. The delta function in the correlator (\ref{noise_def}) can be interpreted in the following sense: when $t_1 \neq t_2$, 
the noise is uncorrelated. In contrast, when $t_1 = t_2 = t$, the variance $\langle \eta^2(t)\rangle = 2 D/\Delta t$.
A realization of the dynamics of such a process is shown in Fig. \ref{walk_picture}. Depending on the parameters
$b$ and $d$, the average number of particles at time $t$ in the system exhibits different asymptotic behaviors.   
For $b < d$, the {\it subcritical} phase, the process dies and on an 
average there are no particles at late times.
For $b > d$, the {\it supercritical} phase, the process is explosive and the average
number of particles grows exponentially with time $t$. In the borderline $b=d$ case, 
the system is critical, where on an average there is exactly one particle in the system at all times.   
This critical case is relevant to several physical and biological systems with stable population distributions \cite{sawyer}. 

BBM is a paradigmatic model of branching processes with wide applications and 
has been studied extensively in both mathematics and physics literature~\cite{fisher,sawyer,harris,bramson,mckean}. 
In one dimension, the positions of the particles at a particular time $t$ represent a set of random variables
that are naturally ordered according to their 
positions on the line with $x_1(t) > x_2(t) > x_3(t) ....$ (see Fig. \ref{walk_picture}). 
It is then interesting 
to study their order statistics, where one is concerned with the distribution of $x_k(t)$, which denotes the position
of the $k$-th rightmost particle. An equally interesting quantity is the spacing between consecutive particles, 
$g_k(t) = x_k(t) - x_{k+1}(t)$ as well as the density of the particles near the tip of the branching process \cite{brunet_derrida_epl,brunet_derrida_jstatphys,ramola_majumdar_schehr}. The questions related to the extremes
in this one-dimensional BBM have been studied extensively over the last few decades \cite{sawyer,bramson,mckean, brunet_derrida_epl,brunet_derrida_jstatphys}. More recently, extreme statistics in this system have found new
applications in the context of estimating the perimeter and area of the convex hull of two-dimensional epidemic spreads \cite{majumdar_pnas}.


Indeed BBM is a useful toy model to study the broader question of extreme value statistics (EVS) of correlated
random variables, a field that has been growing in prominence in recent years.   
Several important properties sensitive to rare events can be characterized by EVS in a wide variety of
disordered systems \cite{bouchaud_mezard, dean_majumdar,paul_satya}. Although probability distributions functions (PDFs) of the extreme values of 
uncorrelated variables are well understood \cite{gumbel},
the computation of extreme and near-extreme value distributions for 
strongly correlated variables constitute important open problems in this field \cite{sabhapandit_majumdar,schehr_majumdar}.
Random walks and Brownian motion have recently proved to be useful laboratories where several exact results
concerning EVS of correlated variables can be obtained~\cite{ramola_majumdar_schehr,schehr_majumdar,perret}. 
In this context BBM represents a useful model where the relevant random variables (the particle positions at time $t$) 
are strongly correlated, and yet exact results concerning the extremes can be obtained. In a recent Letter \cite{ramola_majumdar_schehr}
we briefly discussed some of these results for the critical $b=d$ case. The purpose of the present paper is twofold: (i) to provide
a detailed derivation of these exact results for the critical case and (ii) to extend these results to off-critical cases $b \neq d$.


In the supercritical regime ($b>d$), the statistics of the position of the rightmost particle $x_1(t)$ has been studied for a long time
\cite{bramson,mckean}. In particular, for the case $d=0$,
the cumulative distribution of $x_1(t)$ is known to be governed by the  
Fisher-Kolmogorov-Petrovskii-Piscounov (FKPP) equation \cite{fisher, kpp}. This equation exhibits a travelling front solution: 
the average position of the rightmost particle 
increases linearly with time $\langle x_1(t) \rangle \sim v t$ with a constant velocity $v$ while the width of the front remains
of ${\cal O}(1)$ at late times.
Very recently, Brunet and Derrida studied (still for $d=0$) the order statistics, i.e., the statistics of the positions of the 
second, third, etc $x_2(t), x_3(t) \ldots$. They found that, while $x_k(t) \sim v t$ at late times, with the same speed $v$ for all $k$,  
the distributions of the gaps $g_k(t)$ become independent of $t$ for large $t$, while retaining a non-trivial $k$-dependence \cite{brunet_derrida_epl,brunet_derrida_jstatphys}. They also computed the PDF of the first gap $g_1(t)$ numerically to 
very high precision and also provided an argument for the observed exponentially decaying tail. Several
natural questions remain outstanding. For instance, can one calculate the gap distributions for arbitrary $k$ for $d=0$ as well
as for arbitrary $b$ and $d$?


As mentioned earlier, in a recent Letter, we were able to compute the order and the gap statistics 
of BBM at the critical point $b=d$ at a fixed time $t$, 
by conditioning the process to have a given number of particles at time $t$~\cite{ramola_majumdar_schehr}. As 
we will demonstrate in this paper, this method of conditioning allows us to circumvent the technical difficulties
arising from the inherent non-linearities of the problem and provides exact results for arbitrary $b$ and $d$. 
Let us briefly summarize our main results. Upon conditioning the system to have exactly $n$ particles at time $t$, 
we derive an exact backward Fokker-Planck (BFP) equation for the joint distributions of the ordered positions of the
$n$ particles at time $t$. These equations can, in principle, be solved recursively for all $n$ and the asymptotic results at late times
for any fixed $n$ can be extracted explicitly. We find that at large times, and for all $b$ and $d$, the PDFs of the positions $x_k$'s behave diffusively, 
$P(x_{k},t \to \infty|n) \to \frac{1}{\sqrt{4 \pi D t}}\exp\left(-\frac{{x_{k}}^2}{4 D t}\right)$, with $k = 1,2 É$. Note that for $b > d$, this 
diffusive behavior is in contrast with the case without conditioning on the particle number where it is ballistic. However, as in the case
without conditioning, the PDFs of the gaps $g_k(t)$ become stationary in the long time limit. Moreover we show that the stationary gap PDF has an exponential tail in the super-critical ($b>d$) and sub-critical ($b<d$) regimes and an algebraic tail with exponent $-3$ at the critical point ($b=d$). We argue that these asymptotic tails are {\it universal} in the sense that they are independent of both $n$ (the particle number) and $k$ (the label of the gap). We also discuss the qualitative differences between the conditioned and unconditioned
BBM processes.

The paper is organized as follows. In section 2, we first compute the mean number of particles at time $t$ after which we show in section 3 how to compute the statistics of the rightmost particle using a BFP approach. In section 4, we generalize the BFP approach to compute the (conditional) gap statistics between the two rightmost particles, first in the two-particle sector ($n=2$), and then for an arbitrary number of particles $n \geq 2$. In section 5, we present an asymptotic analysis of the PDF of the first gap for any $n$, which we then generalize to the $k$-th gap. In section 6, we present a comparison of our analytical results with Monte Carlo simulations, before we conclude in section~7.    

\vspace{-0.5cm}
\section{Number of Particles in the System}
\vspace{-0.5cm}

The number of particles $n(t)$ at time $t$ in the one-dimensional BBM 
is a random variable, whose distribution can be computed exactly for 
all $b$ and $d$. Let $P(n,t)$ be the probability that starting with one 
particle at time $t=0$, there are exactly $n$ particles at time $t$. 
One can derive a backward evolution equation for $P(n,t)$ by 
considering all microscopic moves that happen in the initial small time 
interval $\Delta t$. In this small interval $\Delta t$, the particle 
either dies with probability $d \Delta t$, splits into two 
particles with probability $b \Delta t$ and with the remaining 
probability $1-(b+d)\Delta t$ it diffuses. It is easy to see then that

\begin{eqnarray}\label{backward_Pn}
\hspace*{-0.1cm}P(n,t+\Delta t) &=& [1-(b+d)\Delta t] P(n,t) + b \Delta t \sum_{m=0}^n P(m,t) P(n-m,t) \nonumber \\
&+& d \Delta t \, \delta_{n,0} \;.
\end{eqnarray}
By taking the limit $\Delta t \to 0$, this reduces to a partial differential equation
\begin{eqnarray}\label{backward_Pn_cont}
\frac{\partial P(n,t)}{\partial t} = -(b+d) P(n,t) +  b \sum_{m=0}^n P(m,t) P(n-m,t) + d  \, \delta_{n,0} \;.
\end{eqnarray}
This equation (\ref{backward_Pn_cont}) can be solved by a standard generating function technique. One gets the
following explicit solutions:
\begin{eqnarray}
\hspace{-0.5cm}P(0,t) = \frac{d(e^{bt} - e^{dt})}{b e^{bt} - de^{dt}}\;, \hspace{0.5cm}
P(n\ge1,t) = (b-d)^2 e^{(b+d)t} \frac{b^{n-1}(e^{bt}-e^{dt})^{n-1}}{(be^{bt} - de^{dt})^{n+1}}\;.
\label{particle_probabilities}
\end{eqnarray}
The average number of particles in the system at a particular time $t$ is then
\begin{equation}
\langle n(t) \rangle = \sum_{n=1}^{\infty}n P(n,t) = e^{(b-d)t}\;.
\end{equation}
When $b > d$ the number of particles grows exponentially, whereas when $b < d$ the average number of particles
decreases to zero exponentially with time. Exactly at the critical point $b=d$, $\langle n(t)\rangle = 1$ for all $t$. 

Note that, at the critical point, $P(n,t)$ is given by
\begin{eqnarray}
P(0,t)=\frac{bt}{1+bt}, ~~~~~~~~~
P(n\ge1,t)=\frac{(bt)^{n-1}}{(1+bt)^{n+1}} \;.
\label{particle_probabilities_critical}
\end{eqnarray}
Hence, for large $t$, the probability to have $n>0$ particles decays to zero as a power law $P(n>0,t) \sim 1/t^2$ while
the probability of having no particles approaches to unity also as a power law $P(0,t) \sim 1 - 1/(bt)$. In this critical case, although the system
becomes empty of particles almost surely, the average number of particles remains unity at all times. This indicates that
rare events dominate the average behavior and that large fluctuations play a rather important role.

\vspace{-0.5cm}
\section{Statistics of the Rightmost Particle}
\vspace{-0.5cm}
\label{stat_rightmost}
We begin by analysing the behaviour of the rightmost particle in the system at time $t$. 
For this purpose it is convenient to introduce $C(n,x,t)$, denoting the joint probability that 
there are $n$ particles in the system at time $t$, and that all the particles are to the left of $x$. 
The probability $C(0,x,t)$ does not have any clear meaning, but for convenience we choose 
$C(0,x,t) = P(0,t)$. The conditional probability that all the particles lie to the left of $x$,
conditioned on the fact that there are exactly $n$ particles at time $t$ 
is given by $Q(x,t|n) = C(n,x,t)/P(n,t)$, where $P(n,t)$ is given in Eq. (\ref{particle_probabilities}).  
The PDF of the position of the rightmost particle is then given by
$P(x,t|n) = \frac{\partial}{\partial x}Q(x,t|n)$. By definition 
$Q(x,t|n)$ satisfies the boundary conditions $Q(x \to \infty,t|n) = 1$ and $Q(x \to -\infty,t|n) = 0$. 
Initially, since the process starts with a single particle at the origin, it is evident that
$P(n,0) = \delta_{n,1}$ and $C(n,x,0) = \delta_{n,1} \theta(x)$, where $\theta(x)$ is the Heaviside theta function. 
Consequently, the initial condition for the conditional probability is given by $Q(x,0|n) = \theta(x)$ for $n>1$. For $n=0$, 
we recall that $Q(x,0|0) = 1$ by our convention.

\begin{figure*}[h!]
\hspace{1cm}
\includegraphics[width=0.75\linewidth]{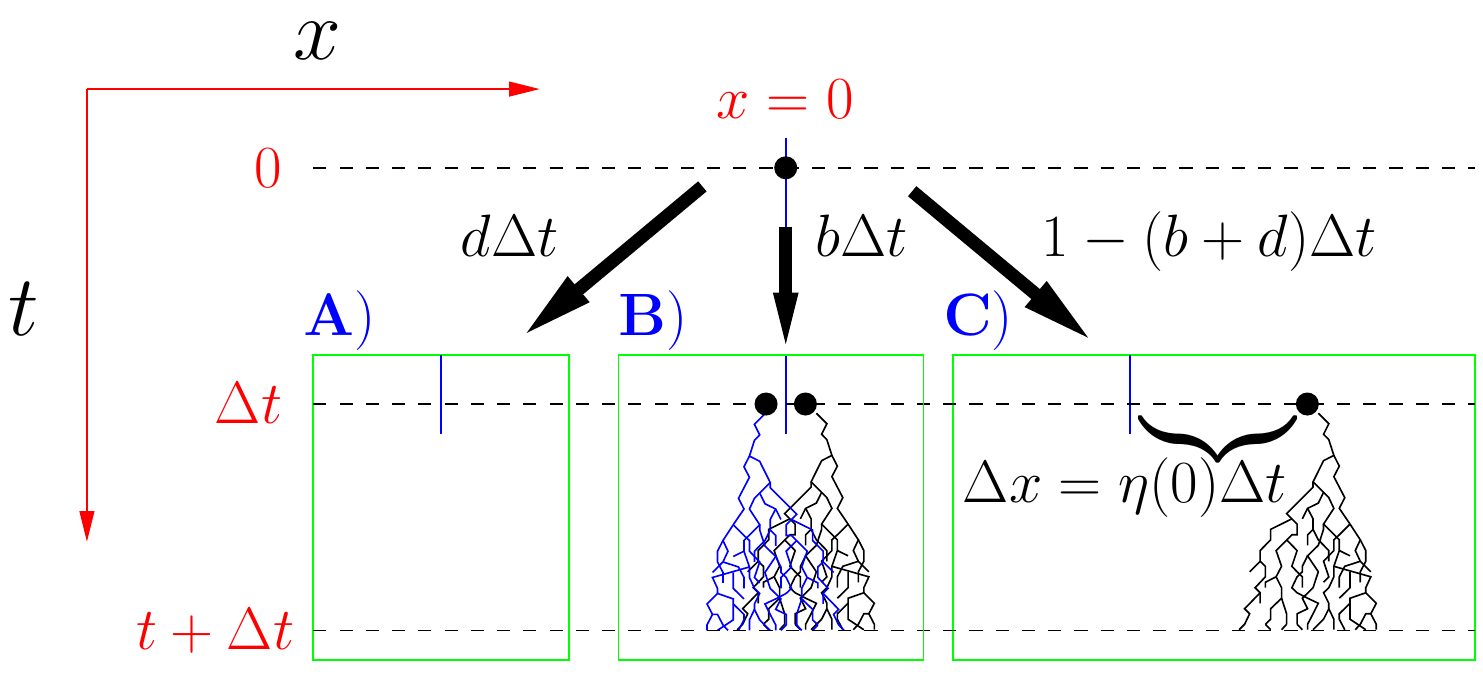} 
\caption{The backward Fokker-Planck approach: In the first time interval  $[0,\Delta t]$,
the particle can {\bf A)} die {\bf B)} split into two independent particles or {\bf C)} diffuse by a distance 
$\Delta x = \eta(0) \Delta t$, 
with probabilities $d \Delta t$, $b \Delta t$ and $1 - (b + d) \Delta t$ respectively. We then
look at the contribution from each of these events to the probabilities at time $t + \Delta t$.}
\label{BFP_fig}
\end{figure*}
\vspace{-0.5cm}
\subsection{Backward Fokker-Planck equation for $C(n,x,t)$}
\vspace{-0.5cm}
\label{bfp_qn_derive}
In this subsection, we start by deriving a BFP equation for the joint probability $C(n,x,t)$. 
To see how $C(n,x,t)$ evolves into $C(n,x,t+\Delta t)$ in a small time interval $\Delta t$, 
we split the time interval $[0,t +\Delta t]$ into two
subintervals: $[0,\Delta t]$ and $[\Delta t, t+\Delta t]$. The system first evolves from its initial condition
to a new configuration at time $\Delta t$ which then acts as a new initial condition for the subsequent evolution
of duration $t$ over the second subinterval $[\Delta t,t+\Delta t]$. 
We next enumerate the probabilities of all the events that take place in the first subinterval $[0,\Delta t]$ (see Fig. \ref{BFP_fig}).
In this subinterval $[0,\Delta t]$, the particle initially at $x=0$:

{\bf A)} dies with probability $d \Delta t$, leading to $n = 0$ particles at all subsequent times.
The contribution to the probability $C(n,x,t+\Delta t)$ from this term is then $d \Delta t \, \delta_{n,0}$.

{\bf B)} splits with probability $b \Delta t$, resulting in two particles at $x =0$. These two particles
give rise to two independent sub-trees. Let $r$ and $n-r$ denote the number of particles in the left and
the right sub-trees respectively, where $0 \leq r \leq n$. Using the independence of the sub-trees, the net contribution
from this event to $C(n,x,t+\Delta t)$ is $b \Delta t \sum_{r = 0}^{n} C(r,x,t) C(n-r,x,t)$.

{\bf C)} diffuses with probability $1-(b+d)\Delta t$, moving a distance $\Delta x = \eta(0)\Delta t$ in the first time step. 
This effectively shifts the entire process by a distance $\Delta x$.
The contribution from this term is then $\left(1- (b+d)\Delta t\right) \langle C(n,x - \eta(0)\Delta t, t) \rangle_{\eta(0)}$. 
Here, and in the following, $\langle \ldots \rangle_{\eta(0)}$ denotes an 
average over all possible values of the diffusive jump at the first time step.

Adding the contributions from terms {\bf A)}, {\bf B)} and {\bf C)}, we arrive at 
\begin{eqnarray}
\nonumber
\hspace*{-0.6cm}C(n,x,t + \Delta t) &=&
\left(1- (b+d)\Delta t\right) \langle C (n,x - \eta(0)\Delta t, t) \rangle_{\eta(0)}\\
\hspace*{-0.6cm}&+&   b \Delta t \sum_{r =0}^{n} C(r,x,t) C(n-r,x,t) + d \Delta t \, \delta_{n,0} \;.
\label{fokker_planck_1}
\end{eqnarray}
Next, using the properties of the Brownian noise in Eq. (\ref{noise_def})
we can Taylor expand Eq. (\ref{fokker_planck_1}) up to second order in $\Delta t$. Taking the limit $\Delta t \to 0$ we arrive at
the backward evolution equation for the cumulative probability
\begin{eqnarray}
\nonumber
&&\frac{\partial C(n,x,t)}{\partial t} = D \frac{\partial^2 C(n,x,t)}{\partial x^2} - (b +d)C(n,x,t) \\
&&\hspace{2.5cm}+  b \sum_{r = 0}^{n} C(r,x,t) C(n-r,x,t) + d \, \delta_{n,0} \;.
\label{FKPP_Q0}
\end{eqnarray}
Using $C(0,x,t) = P(0,t)$ with $P(0,t)$ given in Eq. (\ref{particle_probabilities}), Eq. (\ref{FKPP_Q0}) reduces to
\begin{eqnarray}
&&\frac{\partial C(n,x,t)}{\partial t} = D \frac{\partial^2 C(n,x,t)}{\partial x^2} - (b +d)C(n,x,t) \nonumber \\
&&\hspace{1.8cm}+  2b P(0,t) C(n,x,t) + b \sum_{r = 1}^{n-1} C(r,x,t) C(n-r,x,t) + d \, \delta_{n,0} \;. 
\label{FKPP_Q}
\end{eqnarray}

 If one sums over
the particle number $n$ one gets the cumulative probability distribution of the rightmost particle for the   
{\it unconditioned} BBM: $F(x,t) = \sum_{n=0}^{\infty} C(n,x,t)$. Summing Eq. (\ref{FKPP_Q}) over $n$ one
recovers
\begin{eqnarray}
\frac{\partial F(x,t)}{\partial t} = D \frac{\partial^2
F(x,t)}{\partial x^2} - (b + d)F(x,t) +  b F^2(x,t) + d \;,
\label{FKPP_C}
\end{eqnarray}
together with the boundary conditions $F(x \to +\infty ,t) = 1$ and $F(x \to -\infty ,t) = 0$, for all time $t$. 
For $d>b$ (super-critical phase) the above equation belongs to the FKPP type of non-linear equations \cite{fisher, kpp} which allow
for a traveling front solution at late times $F(x,t) \to F(x-vt)$ with a well defined front velocity $v$ \cite{bramson, mckean}. 
In contrast, for $b=d$ (in the critical case), one can show that the solution of (\ref{FKPP_C}) is diffusive at late times (the non-linearities give rise to only sub-leading corrections). Unfortunately, for finite $t$, this non-linear equation (\ref{FKPP_C}) is not exactly solvable. 
In contrast, by restricting ourselves to a fixed particle number $n$ sector (without summing over $n$) we obtain a set of 
linear equations in $C(n,x,t)$ (\ref{FKPP_Q}). For any given $n$ the terms in the right hand side of Eq. (\ref{FKPP_Q}) involve
the solution $C(m,x,t)$ with $m<n$. Hence, one can solve these linear equations recursively starting from $n=1$, for all $t$ and for all $b$ and $d$. 
That is the trade-off in order to avoid the non-linearities.

\vspace{-0.5cm}
\subsection{Late time behaviour of the conditional probability $Q(x,t|n)$}
\vspace{-0.5cm}

Using Eq. (\ref{FKPP_Q}) for $C(n,x,t)$ and Eq. (\ref{particle_probabilities}) for $P(n,t)$ one can then
write the evolution equation for the conditional probability $Q(x,t|n) = C(n,x,t)/P(n,t)$ explicitly.  
To proceed, it is convenient to first define
 \begin{equation}
f(t) =  2 b P(0,t) - (b+d) = (d-b)\frac{be^{bt}+ de^{dt}}{be^{bt}-de^{dt}} \;.
\label{ft_expression}
\end{equation}
We can then remove the linear term  in Eq. (\ref{FKPP_Q}) by making the transformation
\begin{eqnarray}
C(n,x,t) = e^{\int f(t')dt'} C^{\circ}(n,x,t) = \frac{e^{(b+d)t}}{(b e^{bt}-d e^{dt})^2} C^{\circ}(n,x,t) \;.
\label{Q_transform}
\end{eqnarray}
Inserting this expression into Eq. (\ref{FKPP_Q}), we arrive at
\begin{equation}
\hspace{-0cm}
\frac{\partial C^{\circ}(n,x,t)}{\partial t} = D\frac{\partial^{2}C^{\circ}(n,x,t)}{\partial x^{2}}  
+ \frac{be^{(b+d)t}}{(b e^{bt}-d e^{dt})^2} \sum_{r = 1}^{n-1}C^{\circ}(r,x,t)C^{\circ}(n-r,x,t) \;.
\end{equation}
Next, using Eq. (\ref{Q_transform}) and the expression for $P(n,t)$ in Eq. (\ref{particle_probabilities}) one gets
\begin{equation}\label{eq:Qn_1}
Q(x,t|n) = \frac{C(n,x,t)}{P(n,t)} = \frac{1}{(b - d)^2} \left( \frac{b e^{b t} - d e^{d t}}{b (e^{b t} - e^{d t})} \right)^{n-1}  C^{\circ}(n,x,t).
\end{equation}
The evolution equation for $Q(x,t|n)$ can then be finally written as
\begin{eqnarray}
\nonumber
&&\frac{\partial Q(x,t|n)}{\partial t}  =  
D\frac{\partial^{2}Q(x,t|n)}{\partial x^{2}} + \\
&&\frac{(b-d)^2e^{(b+d)t}}{(e^{b t} - e^{d t})(b e^{b t} - d e^{d t})}
 \sum_{r = 1}^{n-1}\Big[Q(x,t|r)Q(x,t|n-r) - Q(x,t|n)\Big].
 \label{conditioned_Q_eq}
\end{eqnarray}
As we noted before, this is a linear diffusion equation for any $n$ that involves the solutions of $r < n$
as source terms. This set of equations can then be solved recursively to obtain the exact solutions for any $n$.
For example, inserting $n = 1$ in the above equation, we find that $Q(x,t|1)$ obeys the simple diffusion equation
without any source for all $t$, and has the following exact solution
\begin{equation}\label{Qconditioned_n1}
Q(x,t|1) = \frac{1}{2} \mathrm{erfc}\left(\frac{-x}{\sqrt{4 D t}}\right),
\end{equation}
where $\mathrm{erfc}(x)=\frac{2}{\sqrt{\pi}}\,\int_x^{\infty} e^{-u^2}\, du$ is the 
complementary error function.
The corresponding PDF of the position of the particle conditioned on the event $n=1$ at time $t$ is then
\begin{equation}
P(x,t|1) =  \frac{\partial}{\partial x} Q(x,t|1) = \frac{1}{\sqrt{4 \pi D t}}\exp\left(-\frac{x^2}{4 D t}\right).
\end{equation}
We thus find that, for $n=1$, the solution is purely diffusive at {\it all times}. 
In order to analyse the large time behaviour for general $n$ in Eq. (\ref{conditioned_Q_eq}), we note that
the cumulative probability 
is bounded for all $x$ and $t$ ($0 < Q(x,t|n) < 1$).
Therefore, at large $t$, the source term in Eq. (\ref{conditioned_Q_eq}) 
tends to zero as $\sim e^{-|b-d| t}$ (for $b \neq d$), and $\sim 1/(b t^2)$
(for $b = d$).
Thus, at large times $Q(x,t|n)$ obeys the simple diffusion equation for all $n\geq1$
and the solution behaves for large $t$ as
\begin{equation}\label{eq:Qn_2}
Q(x,t|n) \sim \frac{1}{2} \mathrm{erfc}\left(\frac{-x}{\sqrt{4 D t}}\right) \;,
\end{equation}
independently of $n$. 
From this one can deduce that the PDF of the rightmost particle is diffusive at large times.
By symmetry, the leftmost particle also behaves diffusively, and indeed one can show that all the 
particles confined between these two extreme values behave diffusively at large times with
$P(x_k,t|n) \sim \frac{1}{\sqrt{4 \pi D t}} \exp\left(-\frac{{x_k}^2}{4 D t}\right)$ for all $1 \leq k \leq n$.

Let us comment on this result which may seem counter-intuitive at first sight, especially in the super-critical phase. 
As described before, in the super-critical phase ($b>d$), the position of the maximum of BBM has a 
traveling front structure, with the position of the rightmost particle increasing linearly with time $x_1(t) \sim v t$.
The effect of conditioning this process on the number of particles $n$ is thus rather drastic in the super-critical
phase: it slows down the motion of the rightmost particle from ballistic to diffusive. 
This can be understood very simply. Without conditioning, the number of particles typically 
grows exponentially as $e^{(b-d)t}$ in the supercritical regime. Upon conditioning to fix $n$, one picks up
contributions only from atypical diffusive trajectories, out of all the possible trajectories up to time $t$. 
On the other hand, in the critical
case $b=d$, conditioning on a fixed number of particles allows us to correctly 
describe the typical late time behavior of the system \cite{ramola_majumdar_schehr}. 

We note that, although the individual behaviour of the particles is diffusive, 
they are strongly correlated.
In order to understand these correlations, we study the gaps between the successive particles. 
For uncorrelated diffusive 
particles these gaps would also display a diffusive behaviour.
However in BBM, quite remarkably as we show in the next section, 
the PDFs of these gaps become stationary at large times.

\vspace{-0.5cm}
\section{Gap Statistics}
\vspace{-0.5cm}
\label{gap_stat}
We next consider the gap statistics for the conditioned BBM process with $n \geq 2$ particles. 
Let $g_1(t)=x_1(t)-x_2(t)$ denote the gap 
between the two rightmost particles.  
To compute the PDF of $g_1(t)$, we 
study the joint PDF $P(n,x_1, x_2,t)$ that there are exactly $n$ particles ($n\geq 2$) at 
time $t$, with the first particle at position $x_1$ and the second at position $x_2 < x_1$. 
We start with the simplest case $n=2$ which turns out to be already nontrivial. 

\vspace{-0.5cm}
\subsection{Two-particle sector ($n=2$)}
\vspace{-0.5cm}
\label{two_part_sector}

\subsubsection{Backward Fokker-Planck equation for $P(2,x_{1},x_2,t)$}
\label{two_part_BFP_derive}
We first derive the equation governing the temporal evolution of $P(2,x_1,x_2,t)$ using 
a similar BFP approach already discussed in section \ref{bfp_qn_derive}. As before, we split
the interval $[0, t+ \Delta t]$ into two subintervals $[0,\Delta t]$ and $[\Delta t,t+\Delta t]$ (see Fig. \ref{BFP_fig}). 
In the first subinterval $[0,\Delta t]$, the particle at $x=0$:

{\bf A)} dies with probability $d \Delta t$, leading to no particle at subsequent times and thus not contributing to 
the probability $P(2,x_1,x_2,t)$.

\begin{figure*}[h!]
\hspace{2cm}
\includegraphics[width=9cm]{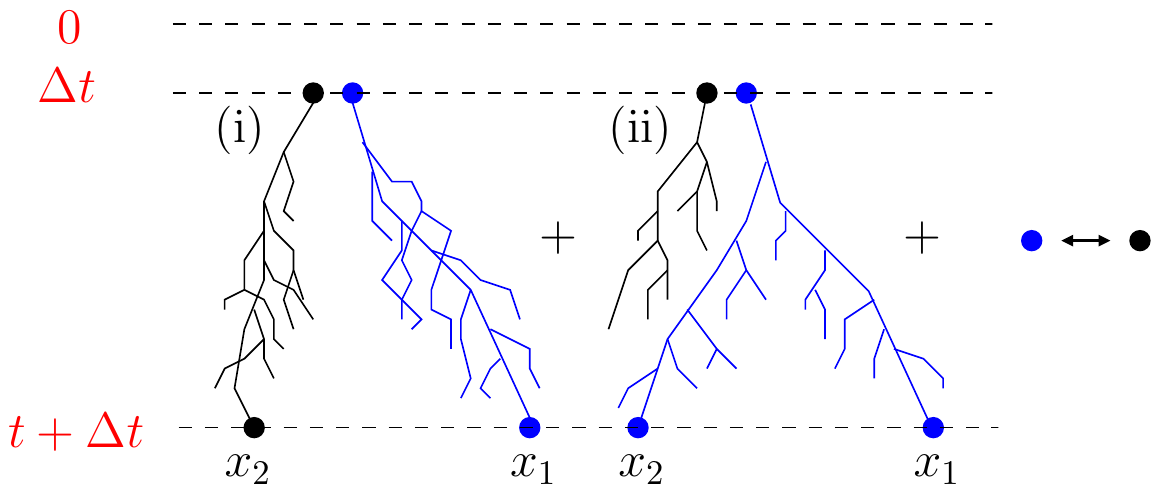} 
\caption{The contribution from the branching term in the BFP equation for the two-particle sector. The particles
at $x_1$ and $x_2$ arise from (i) two different offspring (ii) from the same offspring, generated at the first time step.}
\label{BFP_2_fig}
\end{figure*}

{\bf B)} splits into two particles with probability $b \Delta t$. Here there are two distinct cases to consider (see Fig. \ref{BFP_2_fig}): 
\begin{enumerate}[(i)]
 \item one branch gives rise to a single particle at the final time at position $x_1$ and the other gives rise to a single particle
  at position $x_2$. The contribution from this term is then $2 b \Delta t P(1,x_1,t) P(1,x_2,t)$ where $P(1,x,t)$ is the PDF of having exactly one particle at time $t$ at position $x$. The combinatorial factor $2$ comes from interchanging the two branches. Note that $P(1,x,t) = \partial_x C(1,x,t)$ where $C(1,x,t) = P(1,t) Q(x,t|1)$ with $P(1,t)$ given in 
  Eq. (\ref{particle_probabilities}) and $Q(x,t|1)$ given in Eq. (\ref{Qconditioned_n1}) respectively. This gives explicitly 
\begin{equation}
P(1,x,t) = (b-d)^2 \frac{e^{(b+d)t}}{(b e^{bt}-d e^{dt})^2} \frac{1}{\sqrt{4 \pi D t}}\exp\left(-\frac{x^2}{4 D t}\right) \;.
\label{one_particle_PDF}
\end{equation}

  \vspace*{0.5cm}
  
 \item one branch gives rise to two particles 
  at positions $x_1$ and $x_2$ at the final time and the other gives rise to no particle. 
  The contribution from this term is then $2 b \Delta t P(0,t) P(2;x_1,x_2,t)$.
 \end{enumerate}
 
{\bf C)} diffuses by a distance $\Delta x = \eta(0) \Delta t$
with probability $1-(b+d)\Delta t$. Thus for the second subinterval $[\Delta t,t+\Delta t]$, the process starts from the initial
position $\Delta x = \eta(0) \Delta t$. Hence the contribution from this term is 
$\left(1- (b+d)\Delta t\right) P(2, x_1 - \eta(0) \Delta t, x_2 - \eta(0) \Delta t,t) \rangle_{\eta(0)}$.

Adding the contributions from the terms {\bf A)}, {\bf B)} and {\bf C)} we arrive at
\begin{eqnarray}
\nonumber
&&\hspace*{-1cm}P(2,x_1, x_2, t + \Delta t) = 
\left(1- (b+d)\Delta t\right) \langle P (2,x_1 - \eta(0)\Delta t, x_2 - \eta(0)\Delta t, t) \rangle_{\eta(0)}\\
&&\hspace*{2cm}+2 b \Delta t P(0,t)P(2,x_1,x_2,t) +2 b \Delta t P(1,x_1,t) P(1,x_2,t).
\label{fokker_planck_twopart}
\end{eqnarray}
Expanding the above equation up to second order in $\Delta t$, using the properties of the noise in Eq. (\ref{noise_def})
and taking the limit $\Delta t \to 0$, we arrive at the following evolution equation for the PDF
\begin{eqnarray}
\nonumber
&&\frac{\partial}{\partial t} P(2,x_1,x_2,t) = 
D \left( \frac{\partial}{\partial x_1} + \frac{\partial}{\partial x_2}\right)^2 P(2,x_1,x_2,t)\\
&&\hspace{2cm}+ f(t)P(2,x_1,x_2, t) + 2 b P(1,x_1,t) P(1,x_2,t) \;,
\label{2part_FP}
\end{eqnarray} 
where $f(t)$ is given in Eq. (\ref{ft_expression}).
\subsubsection{Exact solution}
Remarkably Eq. (\ref{2part_FP}) can be solved exactly for all $t$, as we now show. First, it is convenient to get rid of the
second term on the right hand side of Eq. (\ref{2part_FP}) by the customary transformation 
\begin{equation}
P(2, x_1, x_2, t) = \frac{e^{(b+d)t}}{(b e^{bt}-d e^{dt})^2} P^{\circ}(2,x_1, x_2, t).
\label{P2_transform}
\end{equation}
$P^{\circ}(2, x_1, x_2, t)$ then satisfies 
\begin{eqnarray}
\nonumber
&&\frac{\partial}{\partial t} P^{\circ}(2,x_1, x_2, t) = D \left( \frac{\partial}{\partial x_1} + 
\frac{\partial}{\partial x_2}\right)^2 P^{\circ}(2,x_1, x_2, t)\\
&&\hspace{2.5cm}+2b\frac{(b e^{bt}-d e^{dt})^2}{e^{(b+d)t}} P(1,x_1,t) P(1,x_2,t) \;.
\label{P_circ_eqn}
\end{eqnarray}
Next we make the natural change of variables
$s = {(x_1 + x_2)}/{2}$ and $g_1 = x_1 - x_2 > 0$ where $s$ denotes the center of mass and $g_1$ the gap between the
two particles. The Jacobian of this transformation is $1$. The function $P^{\circ}(2,x_1,x_2,t)$ can be expressed as a function of the new coordinates $s$ and $g_1$. In order not to proliferate the number of different functions, we denote this function again by $P^{\circ}(2,s,g_1,t)$ and apologise for this slight abuse of notation. 

Using the explicit expression for $P(1,x,t)$ from Eq. (\ref{one_particle_PDF}) into Eq. (\ref{P_circ_eqn}), we have
\begin{eqnarray}
\nonumber
&&\frac{\partial}{\partial t}{P}^{\circ}(2,s, g_1, t) = D \left( \frac{\partial}{\partial s} \right)^2{P}^{\circ}(2,s, g_1, t)\\
&&\hspace{2cm}+ 2b \frac{ e^{(b+d)t}}{(b e^{bt}-d e^{dt})^2 } (b-d)^4  \frac{1}{4 \pi D t} 
\exp\left(-\frac{2s^2 +\frac{1}{2}g_1^2}{4 D t}\right).
\end{eqnarray}
This is a diffusion equation with a time-dependent source term.
We recall here that the general diffusion equation with a time-dependent source term
\begin{equation}
\frac{\partial}{\partial t} G(x,t) = D \frac{\partial^2}{\partial x^2} G(x,t) + \sigma(x,t),
\end{equation} 
with a given initial condition $G(x,0)$,
can be solved as 
\begin{eqnarray}
\label{diffusion_equation_solution}
\nonumber
&&G(x,t) = 
\int_{-\infty}^{\infty} \frac{dx'}{\sqrt{4 \pi D t}} \exp\left(-\frac{(x-x')^2}{4 D t}\right) G(x',0)\\
&&\hspace{1cm}+\int_{0}^{t}  \frac{dt'}{\sqrt{4 \pi D (t - t')}} \int_{-\infty}^{\infty} dx' 
\exp\left(-\frac{(x-x')^2}{4 D (t-t')}\right) \sigma(x',t') \;.
\end{eqnarray}
Using Eq. (\ref{diffusion_equation_solution})
and the initial condition $P^{\circ}(2,s,g_1,t) = 0$, we arrive at the following exact solution
\begin{eqnarray}
\nonumber
&&P^{\circ}(2,s, g_1, t) = \int_{0}^{t} dt' \int_{-\infty}^{\infty} ds' \frac{1}{\sqrt{4 \pi D (t -t')}} 
\exp\left(-\frac{(s' - s)^2}{4 D (t -t')}\right)  \times\\ 
&& \hspace*{2.5cm} 2b \frac{ e^{(b+d)t'}}{(b e^{bt'}-d e^{dt'})^2 } (b-d)^4 \frac{1}{4 \pi D t'} 
\exp\left(-\frac{2s'^2 +\frac{1}{2}g_1^2}{4 D t'}\right). 
\label{exact_solution_s_g}
\end{eqnarray}
The conditional PDF of the center of mass $s$ and the gap $g_1$, given that there are exactly two particles
in the system at time $t$, is then given by 
\begin{equation}
P(s,g_1,t|2)= \frac{P(2,s,g_1,t)}{P(2,t)}. 
\end{equation}
Using Eq. (\ref{P2_transform}) and the expression for $P(2,t)$ from
Eq. (\ref{particle_probabilities}) we get
\begin{equation}
P(s, g_1, t|2) = \left(\frac{b e^{bt} - d e^{dt}}{b(b-d)^2 (e^{bt}-e^{dt})}\right) P^{\circ}(2,s, g_1, t) \;.
\label{conditioned_two_part}
\end{equation}
Performing the integration with respect to $s'$ in Eq. (\ref{exact_solution_s_g}) and using Eq.~(\ref{conditioned_two_part}) 
we arrive at 
\begin{equation}
{P}(s,g_1,t|2) = \frac{(b-d)^2}{2 \pi D}\left(\frac{b e^{bt} - d e^{dt}}{e^{bt}-e^{dt}}\right) 
\int_{0}^{t} dt' \frac{ e^{(b+d)t'}}{(b e^{bt'}-d e^{dt'})^2 }  \frac{e^{- \frac{g_1^2}{8 D t'}
- \frac{s^2}{2D(2 t- t')}}}{\sqrt{t'(2t-t')}}.
\label{2part_distribution}
\end{equation}
We note that in the limit $d \to b$ this reduces to the expression derived in \cite{ramola_majumdar_schehr}, for the 
gap statistics at the critical point $b =d$, since
\begin{equation}
 (b-d)^2 \frac{ e^{(b+d)t}}{(b e^{bt}-d e^{dt})^2 } \to \frac{1}{(1+bt)^2} ~~~~~\textmd{as} ~~~d \to b \;.
\end{equation}

Given the exact solution of the conditional joint PDF $P(s,g_1,t|2)$ in Eq. (\ref{2part_distribution}) one can
derive the marginal distributions of $s$ and $g_1$ respectively. We start with the center of mass $s$. By integrating over $g_1$ in 
Eq. (\ref{2part_distribution}), we have 
\begin{eqnarray}
\hspace*{-0.5cm}
P(s,t|2) = 
(b-d)^2\left(\frac{b e^{bt} - d e^{dt}}{e^{bt}-e^{dt}}\right) \int_{0}^{t} dt'\frac{ e^{(b+d)t'}}{(b e^{bt'}-d e^{dt'})^2 }
\frac{\exp(- \frac{s^2}{2 D (2 t- t')})}{\sqrt{2 \pi D (2 t- t')}} \;.
\label{com_distribution}
\end{eqnarray}
The integral in (\ref{com_distribution}) is dominated by the region $t' \to 0$, and therefore the marginal PDF of the centre of mass 
behaves diffusively $\sim \frac{1}{\sqrt{4 \pi D t}} \exp \left(-\frac{s^2}{4 D t} \right)$ for large $t$.
This is consistent with the diffusive behaviour of the particles seen in the previous section.
Integrating over the centre of mass variable $s$ in Eq.~(\ref{2part_distribution}), we arrive at the marginal PDF of the gap
\begin{equation}
P(g_1,t|2) = 
(b-d)^2\left(\frac{b e^{bt} - d e^{dt}}{e^{bt}-e^{dt}}\right) \int_{0}^{t} dt'\frac{ e^{(b+d)t'}}{(b e^{bt'}-d e^{dt'})^2 }
\frac{\exp(- \frac{g_1^2}{8 D t'})}{\sqrt{2 \pi D t'}}\;.\\
\label{gap_distribution}
\end{equation}
By taking the limit $d \to b$ in  Eqs. (\ref{com_distribution}) and (\ref{gap_distribution}) 
we recover the expressions derived at the critical point $b=d$ for the marginal PDFs of 
the centre of mass $s$ and the gap $g_1$ respectively, previously obtained in Ref. \cite{ramola_majumdar_schehr}.

For arbitrary values of $b$ and $d$ we find from Eq. (\ref{gap_distribution})
that the gap distribution becomes stationary at large times $P(g_1,t \to \infty|2) = p(g_1|2)$, where the stationary gap distribution
is given by
\begin{equation} 
p(g_1|2) = 
(b-d)^2 \max(b,d) \int_{0}^{\infty} dt'\frac{ e^{(b+d)t'}}{(b e^{bt'}-d e^{dt'})^2 }
\frac{\exp(- \frac{g_1^2}{8 D t'})}{\sqrt{2 \pi D t'}}.\\
\label{asymptotic_distribution}
\end{equation}
Using a saddle point analysis, we can show that the stationary PDF $p(g_1|2)$
has the following asymptotic behaviour for $g_1 \gg 1$
\begin{eqnarray}
p(g_1|2) \sim
\begin{cases}
\dfrac{|b-d|^{3/2}}{\sqrt{2 D} \max(b,d)}\exp\left(-\sqrt{\dfrac{|b - d|}{2 D}} ~g_1\right)\;, ~~~\textmd{for}~~~ b\neq d \;,\\
\\
8\left(\dfrac{D}{b} \right) {g_1^{-3}}\;, ~~~~~~~~~~~~~~~~~~~~~~~~~~~~~~~~~~~~~\textmd{for}~~~ b=d \;. 
\end{cases}
\label{large_g_behaviour}
\end{eqnarray}
It is interesting to note that the expression for the PDF of the gap in the supercritical case $b>d$ turns out to be exponential.
As mentioned above, this behaviour was also obtained for the first gap $g_1 = x_1 - x_2$ in the
{\it unconditioned} BBM \cite{brunet_derrida_jstatphys}. For the case $D = 1$, $b = 1$ and $d =0$, 
the tail was shown to be $\exp(-(1 + \sqrt{2}) g_1)$ for $g_1 \gg 1 $, while in 
the case of the {\it conditioned} process we find from (\ref{large_g_behaviour}) that $p(g_1|2)$ also decays
exponentially albeit with a different rate, namely $p(g_1|2) \sim \exp(- g_1/{\sqrt{2}})$ (see the paragraph after Eq. (\ref{eq:Qn_2}) for a discussion of the origin of the differences between the two processes). It is interesting to note
that the conditioning of the process on $n$ actually {\it decreases} the correlations
between the extreme points, as observing a large gap between the two rightmost particles is more likely in the conditioned process.


\vspace{-0.5cm}
\subsection{$n-$particle sectors with $n>2$}
\vspace{-0.5cm}
\label{n_particle_sector}
When we condition the process to have $n >2$ particles at time $t$, we compute the first gap by 
studying the joint PDF $P(n,x_1,x_2,t)$ that there are exactly $n$ particles in the system at time $t$,
with the first at position $x_1$ and the second at position $x_2 < x_1$. 
Here we also use the BFP approach to derive an evolution equation for this joint PDF. 
The main difference arises in the branching term {\bf B)} at the first time step. For this branching term, and for $n>2$, 
there are three distinct cases to consider (instead of two before):\\
\begin{enumerate}[(i)]
\item One branch gives rise to no particle while the other gives rise to $n$ particles.
The contribution from this term to the final probability is $2 b \Delta t P(0,t) P(n,x_1,x_2,t)$. As noted before in section \ref{stat_rightmost}, the combinatorial factor $2$ comes
from interchanging the two branches.
\item One branch gives rise to $1$ particle while the other gives rise to $n-1$ particles.
The first two particles from the $(n-1)$-particle branch and the particle from the 1-particle branch
are ordered as $x_1 > x_2 > x_3$ at the final time step, with any of them belonging to either branch.
The contribution of this term is
$2 b \Delta t \int_{-\infty}^{x_2} d x_3 \sum_{\tau \in S_3} P(1,x_{\tau_1},t) P(n-1,x_{\tau_2},x_{\tau_3},t)$, where we remind that 
$P(1,x,t)$ is the PDF of having exactly one particle at time $t$ at position $x$, given in Eq. (\ref{one_particle_PDF}).
Here we denote by $\sum_{\tau \in S_N}$ the sum over the permutations $\tau$ of $N$ elements with $\tau_i \equiv \tau(i)$ 
and we use the convention that $P(r,x_i,x_j,t) = 0$ for $i>j$, for any $r \geq 2$.
\item Finally one branch gives rise to $r \ge 2$ particles while the other gives rise to $n-r \ge 2$. The contribution of this term is thus
\begin{equation}
b \Delta t \sum_{r = 2}^{n-2} \int_{-\infty}^{x_2} d x_3 \int_{-\infty}^{x_3} d x_4 \sum_{\tau \in S_4} 
P(r;x_{\tau_1},x_{\tau_3},t)P(n-r;x_{\tau_3},x_{\tau_3},t) \;.
\end{equation}

\end{enumerate}
We can then derive, for any $n > 2$, the BFP equation for $P(n,x_1,x_2,t)$, following the same procedure as explained in section 
 \ref{two_part_BFP_derive} for the case of $n=2$ particles and obtain:
\begin{eqnarray}
\nonumber
&&\frac{\partial P(n,x_{1},x_{2},t)}{\partial t} = D\left( \frac{\partial}{\partial x_1} + \frac{\partial}{\partial x_2}\right)^2 
P(n,x_{1},x_{2},t)+f(t)P(n,x_{1},x_{2},t)\\
&&\hspace{6cm}+b \, \mathcal{S}(n,x_1,x_2,t),
\label{n_particle_diffusion}
\end{eqnarray}
where $f(t)$ is given in Eq. (\ref{ft_expression}) and the source term $\mathcal{S}(n,x_1,x_2,t)$ is obtained by collecting the different contributions computed above:
\begin{eqnarray}
\nonumber
&&\mathcal{S}(n,x_1,x_2,t) = \int_{-\infty}^{x_2} dx_3 \Big[2 \sum_{\tau \in S_3} P(1,x_{\tau_1},t) P(n-1,x_{\tau_2},x_{\tau_3},t) \\
&&\hspace{2cm}+\sum_{r=2}^{n-2}\int_{-\infty}^{x_3} dx_4 \sum_{\tau \in S_4} P(r,x_{\tau_1},x_{\tau_2},t) 
P(n-r,x_{\tau_3},x_{\tau_4},t)\Big] \;,
\label{source_equation}
\end{eqnarray}
where $P(1,x,t)$ is given in Eq. (\ref{one_particle_PDF}). We note that while $x_1$ and $x_2$ stand for the positions of the first and second particle respectively, $x_3$ and $x_4$ are not necessarily the positions of the third and fourth ones.

The BFP equation satisfied by $P(n,x_1,x_2,t)$ (\ref{n_particle_diffusion}, \ref{source_equation})
is a linear diffusion equation for any $n$ that involves the solutions for $P(k,x_1,x_2,t)$ for $k<n$.
Hence, as noted above in section \ref{stat_rightmost}, this set of equations can be solved recursively to obtain the exact solutions for any $n$.
We have computed these expressions for $n = 3$ and $4$, but do not present them here as the expressions are rather cumbersome, being expressible
as a series of nested integrals. One can show that for any $n$, the PDF of the first gap $g_1 = x_1-x_2$ becomes stationary at large times, $P(g_1,t\to\infty|n) \to p(g_1|n)$, which we study below in the large $g_1$ limit. 

\vspace{-0.5cm}
\section{Asymptotic Behaviour}

\begin{figure*}[h]
\includegraphics[width=13cm]{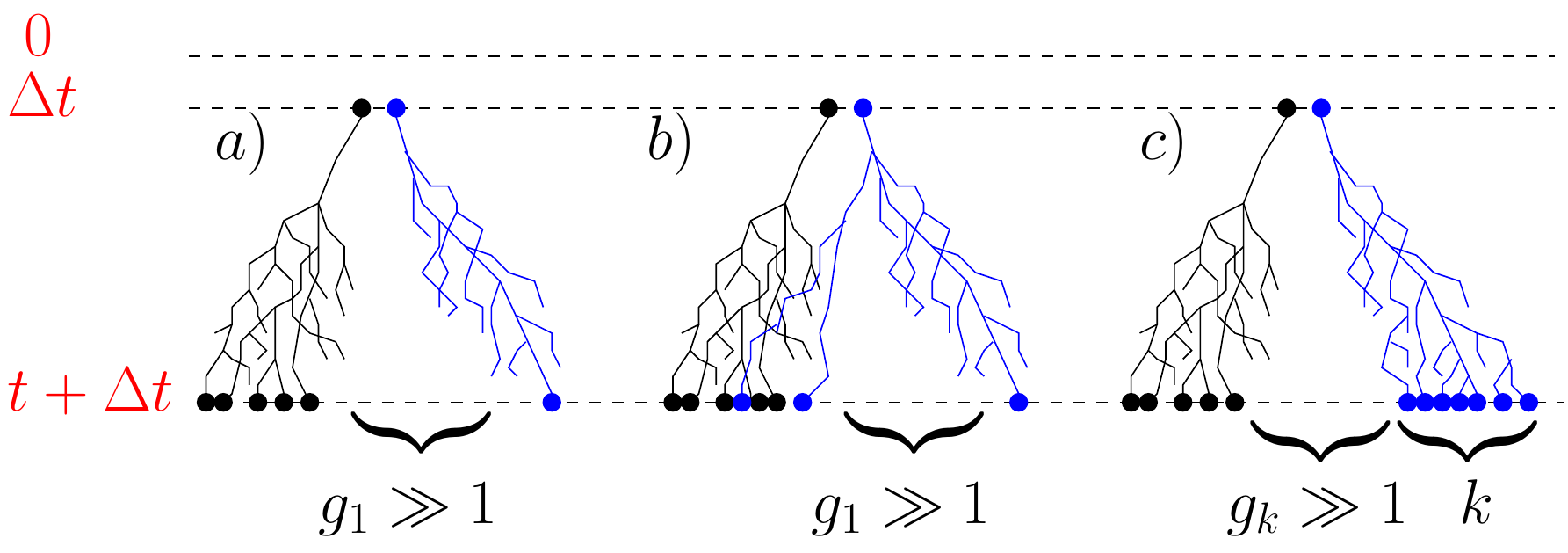} 
\caption{Dominant terms contributing to the large gap behaviour for a) the first gap $g_1(t)$ and c) the $k$-th gap $g_k(t)$. 
Figure b) shows a realization where the large gap is generated by the particles of the same offspring process and is hence suppressed.}
\label{large_gap_fig}
\end{figure*}

Although, the exact expression of the gap distribution $P(g_1,t|n)$ is a bit cumbersome for arbitrary large values of $n$, one can analyze
its large $t$ and large $g_1$ limit, from Eqs. (\ref{n_particle_diffusion}, \ref{source_equation}) as follows. The solution of~(\ref{n_particle_diffusion}) is a linear combination of solutions of individual terms 
in the source function ${\cal S}$ in (\ref{source_equation}). From this, it can be shown that the PDF of the first gap conditioned on $n$ particles
converges to a stationary distribution $P(g_1,t\to \infty|n)=p(g_1|n)$. 
While the full PDF $p(g_1|n)$ in general depends on $n$, 
its tail is independent of $n$.
This can be seen from the fact that the leading contribution to ${\cal S}$ in (\ref{source_equation})
when the gap $g_1 = x_1-x_2 \gg 1$ is large 
arises from the term  in the first line of (\ref{source_equation}) [see Fig. \ref{large_gap_fig} a)]
\begin{eqnarray}
2 b \, P(1,x_1,t) \int_{-\infty}^{x_2} d x_3 P(n-1,x_2,x_3,t) = 2 b \, P(1,x_1,t) P(n-1,x_2,t) \,,
\end{eqnarray}
where $P(n-1,x_2,t) = \partial_{x_2} C(n-1,x_2,t)$ (we recall that $C(n-1,x_2,t)$ denotes the joint probability that 
there are $n-1$ particles in the system at time $t$, and that all the particles are to the left of $x_2$). Since the rightmost particle conditioned on $n-1$ particles in the system behaves 
as a free diffusive particle at large times $P(n-1,x_2,t) \sim P(1,x_2,t)$, see Eqs. (\ref{eq:Qn_1}, \ref{eq:Qn_2}) like in the $n=1$~-~particle case in Eq. (\ref{one_particle_PDF}), we finally obtain that for large $t$
\begin{eqnarray}
2 b \, P(1,x_1,t) \int_{-\infty}^{x_2} d x_3 P(n-1,x_2,x_3,t) \sim 2 b \, P(1,x_1,t) P(1,x_2,t) \,,
\end{eqnarray}
which is precisely the source term for the two-particle case 
analyzed in Eq.~(\ref{2part_FP}). 
This is an advantage of the BFP approach: the two branches arising at the first time step are 
independent of each other at subsequent times.
On the other hand, as we have shown for the two-particle case, the particles from the same branch 
are strongly correlated at large times.
Using this fact, one can show that since all the other terms in ${\cal S}$ in (\ref{source_equation}) involve a larger gap between 
particles generated by the same branch [see Fig. \ref{large_gap_fig} b)], they are suppressed by a 
factor $\int_{g_1}^{\infty} p(g'|k) dg'$, $k<n$ which is exponentially small in the supercritical regime
and falls as a power-law in the critical regime. Therefore, one has that for large $g_1$, $p(g_1|n) \sim p(g_1|2)$ independently of $n\geq 2$, with the asymptotic behaviors given in Eq. (\ref{large_g_behaviour}).

\vspace{0.cm}

Similarly the $k$-th gap $g_k(t) = x_{k}(t) - x_{k+1}(t)$, can be analysed
by studying the joint PDF that 
there are $n$ particles at time $t$ with the $k$-th particle being at position $x_k$ and 
the $(k+1)$-th particle at position $x_{k+1}$. This PDF once again satisfies a 
diffusion equation 
with a source term similar to~(\ref{source_equation}), from which we can show that the PDF of the $k$th gap
reaches a stationary distribution $P(g_k,t\to \infty|n)=p(g_k|n)$. In the large gap limit, 
the dominant term in the source function is the one where the 
first $k$ particles belong to one of the branches generated at the first time step, 
and the subsequent $n-k$ particles belong to the other [see Fig. \ref{large_gap_fig} c)].
This term tends to $2 b P(1,x_k,t) P(1,x_{k+1},t)$ at large $t$, as it involves the leftmost particle
of the first branch being at $x_k$ and the rightmost particle of the other branch being at $x_{k+1}$. 
As noticed before for $g_1$, all other terms involve a large gap between particles generated by the same
branch
and yield subleading contributions when $g_k \to \infty$.  
This implies that {\it the tail of the PDFs of the gaps are universal and are independent of $n$ and $k$}: the large $g_k$ behavior of 
$p(g_k|n)$ is thus given by Eq. (\ref{large_g_behaviour}) with $g_1$ replaced by $g_k$, independently of $n$. 

\vspace{-0.5cm}
\section{Monte Carlo Simulations}
\begin{figure*}[h!]
\hspace{-1cm}
\includegraphics[width=8cm]{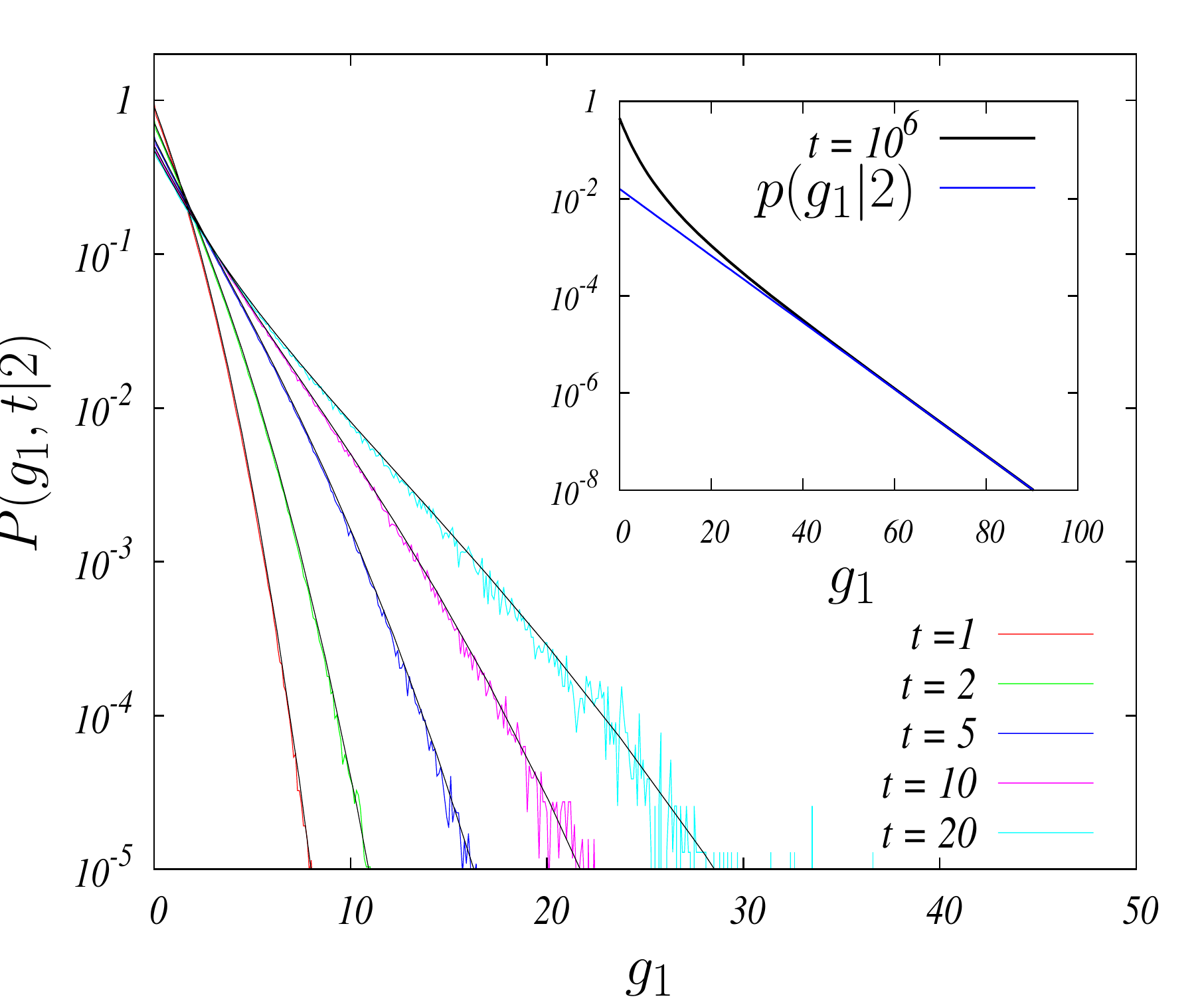}
\includegraphics[width=8cm]{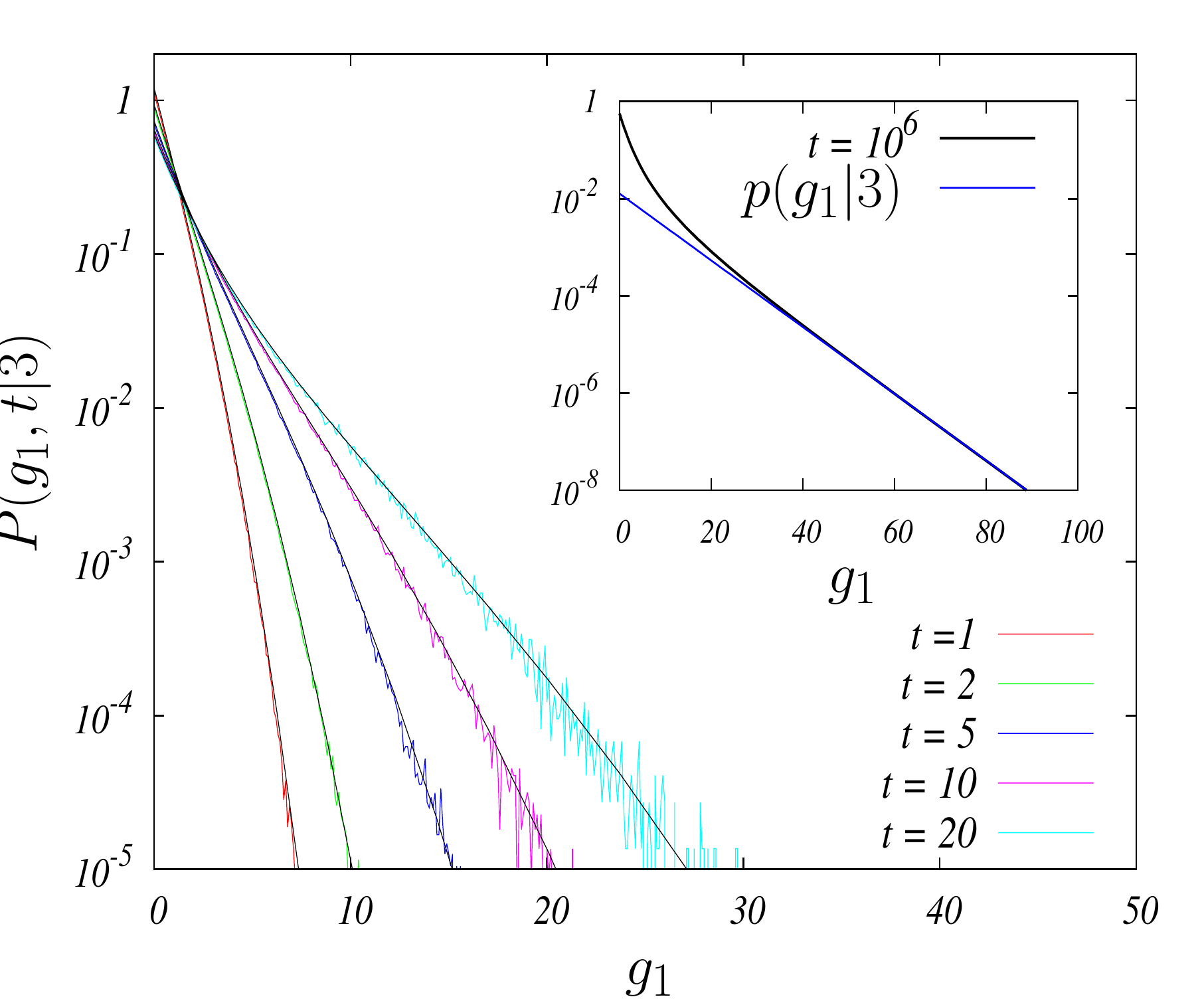} 
\caption{The marginal PDF of the first gap $g_1 = x_1 - x_2$ conditioned on {\bf Left} two particles $P(g_1,t|2)$
and {\bf Right} three particles $P(g_1,t|3)$, at different times obtained from Monte Carlo simulations.
The black lines correspond to the exact theoretical PDFs (given in Eq. (\ref{gap_distribution}) for two particles,
the three particle solution was not given here explicitly as it is rather cumbersome). Here $b = 0.5, d = 0.45$
and $D = 1$. These data have been obtained by averaging over $10^7$ realizations.
In the {\bf Insets} we plot the theoretical PDFs showing the stationary distribution at a late time $t = 10^6$,
along with the
predicted large gap asymptotic behaviour given in Eq. (\ref{large_g_behaviour}).}
\label{2particle_fits}
\end{figure*}

\vspace{-0.5cm}
Finally, we have performed Monte Carlo simulations of the one-dimensional BBM for different values of the parameters $b$ and $d$.
In Fig. \ref{2particle_fits} we plot the marginal PDF of the gap {\it conditioned} on a fixed number $n$ of particles (here $n=2$ and $n=3$). We find a very good agreement between our 
theoretical predictions of the gap PDFs and the distributions extracted from the simulations.

\vspace{-0.5cm}
\section{Conclusion}
\vspace{-0.5cm}
To conclude, we have obtained exact analytical results for the gap statistics of the extreme particles of BBM
conditioned on the number of particles in the system for the general case when $b \ne d$. We derived
backward Fokker-Planck equations governing the distributions of the positions of these extreme 
particles. The conditioning of the PDFs on the number of particles in the system allowed
us to express these evolution equations as a system of linear diffusion equations with source terms, which we
could then solve recursively. We have also obtained exact results for the gap statistics, which can be obtained from
the joint PDF involving the position of two particles. It will be interesting to extend our analysis to the question of $k$-point correlation functions, with $k>2$. 
In this case one can use a similar procedure to analyse the PDF $P(x_1,x_2,x_3....,t|n)$ that given there are exactly
$n$ particles in the system at time $t$, they are at positions $x_1$, $x_2, x_3 ...$. The solutions can in principle 
be obtained in the recursive manner as outlined in our paper.
\ack
SNM and GS acknowledge support by ANR grant
2011-BS04-013-01 WALKMAT and in part by the Indo-French Centre for the Promotion of Advanced Research
under Project 4604-3. GS acknowledges support from Labex-PALM (Project Randmat).


\end{document}